\documentclass[10pt,a4paper,normalheadings,headsepline,headinclude,bibtotoc,fleqn,review]{article}
\usepackage[latin1]{inputenc}
\usepackage{amsmath}
\usepackage{booktabs}
\usepackage{array}
\usepackage{amsthm}
\usepackage{dcolumn}
\usepackage{endnotes}
\usepackage{units}
\usepackage{marvosym}
\usepackage[lines=10]{geometry}
\usepackage{longtable}
\usepackage{rotating}
\usepackage{expdlist}
\usepackage{geometry}
\usepackage{floatrow}
\usepackage{pgfplots}
\pgfmathdeclarefunction{gauss}{3}{% Declare normal distribution function
	\pgfmathparse{1/(#3*sqrt(2*pi))*exp(-((#1-#2)^2)/(2*#3^2))}%
}
\usetikzlibrary{decorations.pathreplacing,calligraphy,backgrounds}
\pgfplotsset{compat=1.16}

\pgfmathsetmacro\valueA{gauss(2.25, 2.25, 0.7)} % Calculate height of the distribution at a specific point

\usepackage{wrapfig}
\usepackage{expdlist}
\usepackage{amssymb}
\usepackage{gloss}
\usepackage{nomencl}
\usepackage{natbib}

\newtheorem{prop}{Proposition}
\newtheorem{coro}{Corollary}

    \makegloss

\renewcommand{\thesection}{\arabic{section}}

\usepackage{fancyhdr} 
\begin{document}

\begin{center}\huge{The Inflation Game}\footnote{I thank Dan Song for able research assistance.} \end{center}

\begin{center} \textit{Wolfgang Kuhle}\\ \textit{Zhejiang University, Hangzhou, China, E-mail wkuhle@gmx.de\\ MEA, Max Planck Institute for Social Law and Social Policy, Munich, Germany\\ VSE, Prague, Czech Republic}\end{center}

\noindent\emph{\textbf{Abstract:} We study a game where households convert paper assets, such as money, into consumption goods, to preempt inflation. The game features a unique equilibrium with high (low) inflation, if money supply is high (low). For intermediate levels of money supply, there exist multiple equilibria with either high or low inflation. Equilibria with moderate inflation, however, do not exist, and can thus not be targeted by a central bank. That is, depending on agents' equilibrium play, money supply is always either too high or too low for moderate inflation. We also show that inflation rates of long-lived goods, such as houses, cars, expensive watches, furniture, or paintings, are a leading indicator for broader, economy wide, inflation.}

\vspace{1cm}
\noindent\textsl{``Markets care until they don't."} Stanley Druckenmiller 

% ----------------------------------------------------
\hspace{.5cm}

\section{Introduction}\label{Introduction}

Economic data for the year 2021, suggest that central bank balance sheets, debt to GDP ratios, as well as the prices of equities, bonds, and real estate, are either at, or near, historic highs. At the same time, inflation rates have remained relatively low during the post crisis years 2009-2020. Against this background, we argue that households may view consumption goods as a new, attractively valued, ``asset class." In turn, we study a simple game, where households discard paper assets in exchange for consumption goods, just to preempt inflation.

Our model is inhabited by plumbers, carpenters, accountants, lawyers, or car mechanics, who do not use general equilibrium theory to think about money and inflation. Instead, our agents consider a one period time horizon, where it is best to buy goods before they increase in price.\footnote{In one interpretation, we study a game, where households choose whether or not to exercise a ``call option" on the goods that are available, at a given sticker price, on the shelves. Once agents have emptied the shelves, prices increase.} Moreover, they take into account that it is costly to buy goods, like a second or third car, just to preempt inflation. In turn, within the context of this simple to understand model, agents play rational expectations equilibria.\footnote{See \citet{Mut61}, \citet{Bla79}, \citet{DeC79}, \citet{Sar93} for a discussion on rational expectations and macroeconomic equilibrium. \citet{Bla89}, Chapter 4, review rational expectations models of money and inflation. As mentioned earlier, our agents are plumbers, carpenters, accountants, lawyers, or car mechanics, who, for better or worse, have never heard of such models.}

%See \citet{Gab20}, \citet{Kuh21} and \citet{Ang21} for recent macro models that depart from rational expectations. In one interpretation, the present simplified model, is in line with theories of rational inattention. Related, we motivated our model with the recent (2009-2021) events surrounding quantitative easing and the increase in asset values.

In equilibrium, inflation rates depend on money supply. If money supply is high, there exists a unique, stable, high-inflation equilibrium. In this equilibrium, households buy additional consumption goods, just to preempt inflation. Moreover, agents' preemptive purchases increase demand and inflation. Finally, high inflation justifies agents' decision to buy additional goods, just to avoid inflation... . If money supply is low, there exists a unique, low-demand, low-inflation equilibrium, where agents have no incentive to buy goods just to avoid inflation. Finally, for intermediate levels of money supply, there exist two stable equilibria, one with high-, and the other with low-inflation. Equilibria with moderate inflation, however, do not exist, and can thus not be targeted. That is, relative to a moderate inflation mandate, central banks can only choose between too much or too little inflation.

We extend our model in Section \ref{Icurve}, where we assume that consumption goods differ regarding their durability. In turn, we show that, as money supply grows, inflation shows up in long-lived goods, such as houses, furniture, cars, or luxury watches, first. Finally, as money supply increases further, a broad basket of shorter-lived consumption goods experiences a discontinuous increase in inflation. Inflation rates of long-lived goods and assets are thus leading indicators for broader increases in inflation.

\section{Deterministic Model}\label{Deterministic}

There is a large number of households $i\in[0,1]$. Each household is endowed with one unit of ``disposable money."\footnote{We may think of ``disposable money" as money, which a household holds even though he does not strictly need it to perform his normal day to day transactions. Such holdings are particularly large in a liquidity trap. Put differently, we may think of a household, who holds paper money, or other paper assets, such as short-term bonds, mainly due to a lack of attractive alternatives, i.e. as a store of value.} Each household $i$ can choose between two actions $a_i=\{1,0\}.$ Households who choose $a_i=1$ discard their money holdings to ``beat inflation." That is, households who choose $a_i=1$, buy goods today, in order to avoid the cost of inflation $\pi$. Households who discard their money fore-go the liquidity service of their money holdings. We denote this cost by $r$. In an alternative interpretation, $r$ may represent the cost of buying a good, like a second car, just to avoid inflation.\footnote{In the context of paper assets like short-term bonds, $r$ may also represent an interest payment. Finally, we note that our results still hold if we assumed an exogenous rate $r=0$, i.e. if we, or the households, disregarded these potential costs.} Households who choose $a_i=0$ keep their money holdings unchanged, and thus earn a net return of $r-\pi$. 

Regarding inflation $\pi$, which may be negative, we assume that it is a continuously differentiable function of aggregate incremental household demand $A=\int_0^1a_idi$ and money supply $M$ such that 
\begin{eqnarray} \pi=\pi(A,M),\quad \frac{\partial \pi}{\partial A}=\pi_A>0, \quad \frac{\partial \pi}{\partial M}=\pi_M>0. \label{Inflation} \end{eqnarray} 
Where $\pi(0,M)$ is the inflation rate that obtains when households do not buy assets to preempt inflation. On the contrary, $\pi(A,M), A>0$ represents a situation where a mass $A>0$ of agents buy additional goods today, just to avoid future price increases. The nominal interest rate on paper assets $r(M)$, which may also be negative, is assumed to be a continuously differentiable, decreasing, function of aggregate money supply. Finally, we assume that $\lim_{M\rightarrow\infty}(r(M)-\pi(1,M))<0$ and $\lim_{M\rightarrow 0}(r(M)-\pi(1,M))>0$. The last two assumptions ensure that (i) households will eventually use money mainly for transaction as $M\rightarrow 0$ and (ii) that discarding money holdings to preempt inflation will eventually be a dominant strategy as $M\rightarrow\infty$.    

Households maximize:
\begin{equation} \max_{a_i}U(a_i)=(1-a_i)(r-\pi)+a_i(\pi-r).  \end{equation} 
Accordingly,
\begin{equation}\label{FOC}
a_i= \left\{\begin{array}{ll}
1 \qquad\qquad & \pi>r \\
0 \qquad\qquad & \pi<r
\end{array} \right.
\end{equation}
Using (\ref{FOC}), we can define $M^*$ and $M^{**}$ such that 
\begin{equation} \pi(A=1,M^*)=r(M^*),  \label{M}\end{equation}
\begin{equation} \pi(A=0,M^{**})=r(M^{**}).  \label{MM}\end{equation}
It follows from our assumptions on $\pi(,)$ and $r()$ that $M^*<M^{**}.$ In turn, we have 
\begin{prop} There exists a unique equilibrium where $A=0$ as long as $M<M^*$. For $M\in[M^*,M^{**}],$ we have two stable equilibria $A=0$ and $A=1$. Finally for $M>M^{**},$ we have a unique equilibrium $A=1$. \label{P1}\end{prop} 
Diagram 1 illustrates Proposition (\ref{P1}), respectively, the game's equilibria. There are two Corollaries to Proposition \ref{P1}:
\begin{coro}The equilibrium inflation rate $\pi(A(M),M)$ is, in one or more points, discontinuous in $M$.\label{C1} \end{coro}
Moreover, if the curve $\pi(A=0,M)$ is flat,\footnote{A liquidity trap is typically associated with flat curves $r(M)$ and $\pi(0,M)$.} there exists an intermediate range of inflation rates, which are neither consistent with the low inflation equilibrium $\pi(A=0,M)$ nor with the high inflation equilibrium $\pi(A=1,M)$:
\begin{coro} If $\pi(0,M^{**})<\pi(1,M^*),$ then there exists no money supply $M$ such that equilibrium inflation $\pi(A(M),M)$ falls into the interval $(\pi(0,M^{**}),\pi(1,M^*)).$ \label{C2}\end{coro}
According to Corollary \ref{C2}, money supply is, depending on agents' equilibrium play, always either too low or too high for moderate inflation. Moderate inflation rates, $\pi(A(M),M)\approx 2\%$, which central banks are supposed to target, may thus not be in the set of possible equilibrium outcomes.\footnote{We argue that the range $(\pi(0,M^{**}),\pi(1,M^*))$ is indeed likely quite large. That is, a change in behavior from $A=0$ to $A=1$, will likely have a big impact on inflation rates. In the context of the US economy, $M1$ increased more that 10 fold (from 1.4 trillion in 2007 to 18 trillion in mid 2020) while nominal GDP increased only 50\% (from 14.5 trillion in 2007 to 21 trillion in 2020). This suggests that a coordinated move, to an $A=1$ equilibrium, in which households discard their (relative to 2007) excessive paper assets, should result in a, using e.g. the quantity equation $PY=Mv$, potentially drastic increase in prices. That is, ceteris paribus, inflation in a $A=1$ equilibrium could easily be in the double or triple digits. At the same time, inflation over the last decade, which we interpret as a $A=0$ equilibrium, was very low. Taking both arguments together, the interval $(\pi(0,M^{**}),\pi(1,M^*))$ may be quite large. More generally, this interval will be large in situations where the ratio between the nominal value of all paper assets on the one hand, and the nominal value of the goods and services that the economy can produce, now or in the future, has grown extremely large.}  

%stocks have gone up spy 150-t 450

\begin{tikzpicture}\label{Dia1}
		\draw[->](-5,0)--(1,0) node [right] {$M$}; % set the x axis
		\draw[->](-5,0)--(-5,4) node [left] {$A$}; % set the y axis
		\draw[dotted,black,thick](-3,2)--(-3,0);
		\node[black,thick] at (-3,-0.3) {$M^*$};
		\draw[dotted,black, thick](-5,2)--(-3,2);
		\node at(-5.3,2) {$1$};
		\draw[black, very thick](-3,2)--(1,2) node[above]{$A(M)$};
		\draw[dotted,black, thick](-1,2)--(-1,0);
		\node[black] at (-1,-0.3) {$M^{**}$};
		\draw[black,very thick](-5,0)--(-1,0);
		\draw [fill,black] (-3,2) circle [radius=.05];
		\draw [fill,black] (-1,2) circle [radius=.05];
		\draw [fill,black] (-3,0) circle [radius=.05];
		\draw [fill,black] (-1,0) circle [radius=.05];
		
		%pictre2
		\draw[->](3,0)--(9,0)node [right] {$M$}; % set the x axis
		\draw[->](3,0)--(3,4)node [left] {$\pi$}; % set the y axis
		\draw[black,very thick](3,0.5)--(7,1.1) node[right]{$\pi(A=0,M)$};
		\draw[black,very thick](5,2.5)--(9,3.1) node[above]{$\pi(A=1,M)$};
		\draw [fill,black] (5,2.5) circle [radius=.05];
		\draw[dotted,black,thick](5,2.5)--(5,0) node [below]{$M^*$};
		\draw[dotted,black,thick](7,2.83)--(7,0) node [below]{$M^{**}$};
		\draw [fill,black] (7,2.8) circle [radius=.05];
		\draw [fill,black] (7,1.1) circle [radius=.05];
		\draw[dotted,black,thick](5,2.5)--(6.95,2.5);		
		\draw[dotted,black,thick](5.1,1.1)--(7,1.1);
		\draw[pen colour={black},decorate, 
		decoration = {calligraphic brace,raise = 2pt, amplitude = 4pt,mirror}](5,2.5) --  (5,1.1);
		\end{tikzpicture}
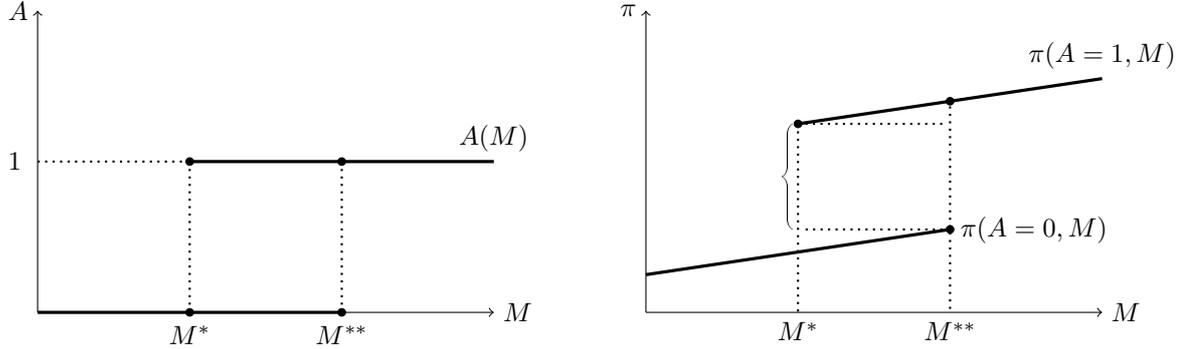
		\begin{center}\textbf{Diagram 1} Equilibrium correspondence and inflation rates. The winged bracket (right diagram) contains intermediate inflation rates, which cannot be targeted. \end{center} 

\subsection{An Empirical Implication: The Inflation Curve}\label{Icurve}

In one interpretation, variable $r$ represents the cost of buying goods before they are needed, just to preempt inflation. This cost is arguably lower for durable consumption goods, such as houses, furniture, or cars, than for less durable/hard to store consumption goods, such as vegetables, a haircut, or a warm meal. We may thus think of an extended model where different goods $j=1,2,3...N$ are ordered by their longevity, respectively, such that $r_{j+1}>r_j$. Moreover, in addition to a lower cost $r_j$, the short-run elasticity of supply will arguable be smaller for hard to make, capital intensive, durable goods, such as cars and houses. That is, we may also/alternatively assume $\pi_{j+1}(A,M)>\pi_j(A,M)\forall A>0$, such that a sudden increase in demand will lead to a greater price movement in long-lived goods with inelastic short-run supply.

Following this argument, we obtain, for each consumption good $j$, analogs to the households' first order condition (\ref{FOC}). Moreover, recalling (\ref{M}) and (\ref{MM}), yields cutoffs $M_{j}^{*}$ and $M_{j}^{**}$, which separate the quantities of money for which equilibrium inflation in market $j$ is either low, indeterminate, or high. Given our assumptions on $r_j$ and/or $\pi_j$, these quantities differ across goods, such that $M_{j+1}^{*}<M_{j}^{*}$ and $M_{j+1}^{**}<M_{j}^{**}$. That is, starting in a stable, low-inflation equilibrium, more and more goods categories, beginning with the most long-lived goods, start to experience a discontinuous uptick in inflation as money supply grows past the respective thresholds $M^{**}_j$. In this interpretation of the model, a sudden uptick in house prices would foretell a subsequent rise in the prices of cars and luxury watches. More generally, the price movements in long-lived goods\footnote{We could extend the current argument, and include the most long lived assets such as stocks, which, depending on their price, should be associated with a negative (positive) cost $r_j<0$ $(r_j>0)$. In this interpretation, stocks would be the first asset class/good, that moves to an equilibrium with high (asset price) inflation. In turn, long lived consumption goods follow. Finally, as agents run out of long lived investment opportunities, inflation reaches non-durable consumption goods.} would be leading indicators for a broad, economy wide, inflation. 

\section{Conclusion}
We study an economy where agents exchange paper-assets for consumption goods, to preempt inflation. In one interpretation, our model mimics the perspective of real-world households, who try to buy goods before they appreciate in price. In turn, if money supply is high, there exists a unique high-inflation equilibrium, where agents buy additional goods, just to avoid the high cost of inflation. Moreover, if money supply is low, there exists a unique low-inflation equilibrium, where agents' demand for additional goods, and thus inflation, is low. 

For intermediate levels of money supply, there exist multiple stable equilibria where inflation is either high or low. It follows that, depending on agents' equilibrium play, money supply is always either too high or too low for moderate inflation. If moderate inflation is a socially desirable outcome, the current model thus predicts that central banks can only choose between too much and too little inflation. Finally, our model also predicts that inflation rates of long-lived goods are a leading indicator for economy wide increases in inflation. 

\newpage

\addcontentsline{toc}{section}{References}
\markboth{References}{References}
\bibliographystyle{apalike}
\bibliography{References}

\end{document}